# Quantum phase transitions in highly crystalline two-dimensional superconductors

Yu Saito[1]*, Tsutomu Nojima[2], and Yoshihiro Iwasa[1,3]

[1] *Quantum-Phase Electronics Center (QPEC) and Department of Applied Physics, The University of Tokyo, Tokyo 113-8656, Japan*
[2] *Institute for Materials Research, Tohoku University, Sendai 980-8577, Japan*
[3] *RIKEN Center for Emergent Matter Science (CEMS), Wako 351-0198, Japan*
*Corresponding author: saito@mp.t.u-tokyo.ac.jp

**Superconductor-insulator transition is one of the remarkable phenomena driven by quantum fluctuation in two-dimensional (2D) systems. Such a quantum phase transition (QPT) was investigated predominantly on highly disordered thin films with amorphous or granular structures using scaling law with constant exponents. Here, we provide a totally different view of QPT in highly crystalline 2D superconductors. According to the magneto-transport measurements in 2D superconducting ZrNCl and $MoS_2$, we found that the quantum metallic state commonly observed at low magnetic fields is converted via the quantum Griffiths state to the weakly localized metal at high magnetic fields. The scaling behavior, characterized by the diverging dynamical critical exponent (Griffiths singularity), indicates that the quantum fluctuation manifests itself as superconducting puddles, in marked contrast with the thermal fluctuation. We suggest that an evolution from the quantum metallic to the quantum Griffiths state is generic nature in highly crystalline 2D superconductors with weak pinning potentials.**

**Introduction**

Two-dimensional (2D) superconductors have been well-known platforms for the study of a quantum phase transition (QPT). In granular or amorphous superconducting thin films, which was the initial 2D superconductors with the atomic layer thickness, a direct superconductor-insulator transition (SIT) was a consequence of the QPT occurring at zero temperature. The SIT has been long discussed based on so called "dirty-boson model", which is applied to the highly and homogeneously disordered systems[1,2]. In this scenario, the ground state is either superconductor or insulator distinguished by a quantum critical point (QCP), which is characterized by the critical sheet resistance and a critical tuning parameter for degree of disorder, magnetic field or carrier density. However, in recently-found highly crystalline 2D superconductors[3], such as the gate-induced single crystal surfaces[4] and the mechanically exfoliated atomic layers[5], the dirty-boson scenario for the magnetic field induced SIT needs to be modified using a different picture incorporating a wide quantum metallic state/vortex liquid state. Such a metallic state following the fragile zero-resistance state occurs because of the quantum creep motion of vortices originating from the minimal disorder and the extremely weak pinning.

Although the resistance in the quantum metallic state is nonzero, this state starts far below the mean field upper critical field ($B_{c2}^{MF}$) without showing localized behavior. Thus, the crossover or transition from the quantum metallic state to the normal state at low temperatures including superconducting fluctuation regime, emerges as a new unsolved problem of the QPTs in the highly crystalline 2D superconductors[6–10,3]. Recently, there appeared a couple of clues for this issue. One is the gate-induced superconducting surface of a ZrNCl single crystal in which the finite size scaling (FSS) law of transport property does not obey the dirty-boson scheme any more[4]. Others are molecular-beam-epitaxy grown Ga trilayer[11], NbSe$_2$ monolayer[12] and LaAlO$_3$/SrTiO$_3$ (110) interface[13] in which a different type of the QCP, the quantum Griffiths singularity[14], was observed at ultra-low temperatures.

Here we propose a comprehensive quantum phase diagram including superconducting fluctuation outside the mean field phase boundary of highly crystalline 2D superconductors. Based on the magneto-transport study in ion-gated ZrNCl and MoS$_2$, we found that these 2D superconducting systems commonly show the wide range of a quantum phase stemming from the very low magnetic field to a QCP beyond $B_{c2}^{MF}$. Especially, the observed QPT is characterized by the diverging behavior of critical exponent toward a QCP known as quantum Griffiths singularity. This implies that the critical phenomenon below the QCP is described by the exponentially small but nonzero probability of large ordered regions, so called rare regions. Such rare regions are interpreted as the superconducting puddles surviving in the normal state background with a long time and length scale at $T \to 0$ K[15], which evolves from the short-range ordered vortex state accompanied by the quantum vortex creep at low magnetic fields below $B_{c2}^{MF}$. The present observation strongly indicates that quantum fluctuation governs the wide quantum metallic state and its crossover to the quantum Griffiths state consisting of superconducting puddles, leading to a generic feature of highly crystalline 2D superconductors.

**Results**

**Superconductivity and quantum metal in ion-gated ZrNCl and MoS$_2$**    ZrNCl and MoS$_2$ are originally band insulators with honeycomb layered crystal structures (Fig. 1a), both of which exhibit superconductivity by alkali–metal intercalation or field-effect doping[3]. We prepared electric-double-layer transistors (EDLTs) with those single-crystal flakes (see methods for details), in which almost all the carriers are accumulated in the topmost layer with 1~2 nm at high gate voltages, resulting in truly 2D superconductivity at low temperatures[4,16].

Figure 1b displays temperature-dependent sheet resistance $R_{sheet}(T)$ at $V_G$ = 6.0 V (sheet carrier density $n_{2D}$ is $4.8 \times 10^{14}$ cm$^{-2}$ at 30 K) for zero and finite magnetic fields $B$ applied perpendicular to the surface of ion-gated ZrNCl. In zero magnetic field, the ZrNCl-EDLT exhibits superconductivity with a critical temperature $T_c$ of 14.6 K. Here, $T_c$ is defined as the temperature where the sheet resistance becomes a half of the normal state value, and is almost equal to the mean-field transition temperature $T_{c0}$ estimated using thermal fluctuation theories[4]. By applying $B$, the resistive transitions considerably broadened because of the thermal phase fluctuation, and then level off at low temperatures due to the quantum phase fluctuation, indicating the quantum metallic state. This quantum metallic state, which appears once the magnetic field is switched on, possibly originates from weak pinning potentials and large fluctuations reflecting the two-dimensionality[4]. Such a stable quantum metallic state does not appear in atomically ordered 2D superconductors grown by molecular-beam-epitaxy[12,17,18], possibly because of the atomic steps of substrate working as strong pinning centers. In the exfoliated 2D superconductors like ZrNCl, MoS$_2$ and NbSe$_2$, such atomic steps are basically absent[4,5,16].

**Estimation of mean-field upper critical field $B_{c2}^{MF}$**  For establishing a global phase diagram of 2D superconductors, the estimation of both mean-field upper critical field $B_{c2}^{MF}(T)$ and superconducting onset are crucial. For this purpose, we adopted the Ullah-Dorsey (UD) scaling theory[19] (see Supplementary Note 1 for details) to calculate $B_{c2}^{MF}(T)$, where the excess conductance due to the thermal superconducting fluctuation, $G_{fl} \equiv 1/R_{sheet}(T) - 1/R_N(T)$, in finite $B$ is described with the Hatree approximation as follows:

$$G_{fl}\left(\frac{B}{T}\right)^{1/2} = F\left(\frac{T - T_c^{MF}(B)}{(TB)^{1/2}}\right) \quad F(x) \propto \begin{cases} -x & (x \ll 0) \\ x^{-s} & (x \gg 0) \end{cases}. \quad (1)$$

Here, $R_N(T)$ is the sheet resistance in the normal state, $s = 1$ in the case of a 2D system[20] and $T_c^{MF}(B)$ the mean-field transition temperature in a magnetic field, corresponds to the temperature on the $B_{c2}^{MF}(T)$ curve. As shown in Fig. 1c, by using $R_{sheet}(T)$ at $B = 9.0$ T as $R_N(T)$ and $T_c^{MF}(B)$ as fitting parameters, $G_{fl}(T)$ curves in the ZrNCl-EDLT at $T > T_c^{MF}(B)$ and $B = 0.1 - 1.8$ T collapse onto a single curve with the slope of -1 in log-log plots, obeying the UD scaling law of eq. (1). Then we plotted $T_c^{MF}(B)$ or $B_{c2}^{MF}(T)$ as shown in Fig. 1d together with the superconducting onset temperature and field, $T_{onset}$ and $B_c$, which are defined later. The $T_c^{MF}(B)$ at 0.1 T is close to the temperature for 50 % of the normal state resistance. With increasing $B$, $R_{sheet}(T_c^{MF})/R_N$ increases and goes over 90% around $T_c^{MF}(B) \sim T_{c0}/2$. At lower temperatures below $T_{c0}/2$, where the UD scaling theory does not holds[19], we simply defined $T_c^{MF}(B)$ as the temperature for 95 % of the normal resistance. $T_c^{MF}(B)$ or $B_{c2}^{MF}(T)$ thus obtained obeys the Wertharmer-Helfand-Hohenberg (WHH) theory[21] as shown in the dashed curve in Fig. 1d. By using the similar process as the ZrNCl-EDLT, we determined $T_c^{MF}(B)$ or $B_{c2}^{MF}(T)$ curve of a MoS$_2$-EDLT ($V_G = 5.0$ V and $n_{2D} = 1.3 \times 10^{14}$ cm$^{-2}$) as shown in Fig. 1e.

**Crossover from quantum metal to quantum Griffiths state**   After knowing the mean field curve, we next investigate the superconducting onset. Figures 2a – d show the magnetoresistance at various temperatures in ZrNCl and MoS$_2$. In both cases, the isotherms show multiple crossing points, indicating that each $R$-$B$ curve cross at different points. The cross points of $R$-$B$ curves at neighboring temperatures, which are defined as $B_c$(T), are plotted in Figs. 1d and e as orange squares. In these figures, we also plot the onset temperature $T_{onset}$ of resistance drop in a magnetic field, defined by the temperature where $dR_{sheet}/dT = 0$ (see Supplementary Note 2 for details). Interestingly, $T_{onset}$ traces almost the same curve as that of $B_c(T)$. Because both $T_{onset}(B)$ or $B_c(T)$ curve is above the mean field curve, it is interpreted as the onset temperature (or magnetic field) of superconducting fluctuation. Both Figs. 1d and e represent the large fluctuation region outside the mean field curve, reflecting that the systems are at the 2D limit. More importantly, the interval between $B_c(T)$ and $B_{c2}^{MF}(T)$ increases with decreasing $T$ in ZrNCl or almost unchanged in MoS$_2$, respectively. This behavior cannot be explained by the effect of thermal fluctuation, which results in the decrease of the interval, and signifies the occurrence of the quantum fluctuation at low temperature and high magnetic fields.

The multi-crossing behavior at low temperatures in Figs. 2a-d shows marked contrast to the single crossing point observed in amorphous metallic films, e.g., MoGe[22] and Ta films[23], suggesting that the present system is not classified into the conventional universality class, which exhibits a direct superconductor-insulator transition with a single crossing point. A possible scenario for the quantum fluctuation above $B_{c2}^{MF}(T)$ is the quantum Griffiths singularity, which was first theoretically proposed to understand the divergence of the dynamical exponent in QPTs in various lattice models with randomness[24,25]. The existence of the quantum Griffiths singularity has been experimentally applied first to magnetic metal systems[26–28] and recently to 2D superconductors[11-13] (see Supplementary Note 3 for details).

We have tested the scenario of the quantum Griffiths singularity by introducing the FSS law for magnetoresistance expressed as,

$$R(B,T) = R_c f\left(\frac{B - B_c}{T^{\frac{1}{zv}}}\right) \qquad (2)$$

in various temperature regimes around the crossing points (see Supplementary Note 4 for details). Here, $R_c$, and $B_c$ are the asymptotic critical sheet resistance and the critical magnetic field defined as the values at the crossing points, $f(x)$ the scaling function with $f(0) = 1$, $z$ and $v$ the dynamical and static critical exponent, respectively. To examine the evolution of the effective critical exponents $zv$ with the decrease (or increase) in $T$ (or $B$), five representative crossing points were selected both for ZrNCl and MoS$_2$, respectively, and then the FSS analysis at various temperature regimes were performed. During the scaling, we derived $zv$ by assuming that the functional forms of $f(x)$ both for $x > 0$ and $x < 0$ unchanged (see Supplementary Note 4). As shown in Figs. 3a and b, $z\nu$ is not constant but varies as a function of magnetic field in both systems: in the low magnetic field (high temperature) regime $z\nu$ is around 0.5, while it increases with increasing (decreasing) magnetic field (temperature), exceeding 1 (Figs. 3a and b). Finally, it seems to diverge toward a certain critical field following the relation of $z\nu \sim (B - B_c^*)^{-0.6}$, which is the evidence of quantum Griffiths singularity at characteristic critical magnetic field $B_c^*$. This diverging behavior of the critical exponent with multiple crossing points is in stark contrast to the conventional dirty metallic films, which shows temperature-independent constant critical exponents with a single critical field[2]. Such a divergent critical exponent consistent with the Griffiths singularity indicates that rare ordered regions are existing even above $B_{c2}^{MF}$ with a rather long time-scale and finite length scale due to the quantum fluctuation. Such rare regions are interpreted as superconducting puddles in this particular case.

**Discussion**

Based on the experimental data of the ZrNCl-EDLT, we provide schematic images of quantum Griffiths state (Fig. 4a) and quantum metal (Fig. 4b) and propose a comprehensive phase diagram of highly crystalline 2D superconductors (Fig. 4c), which is characterized by two crossover curves, $B_c(T)$ (orange squares) and $T_{cross}(B)$ (blue triangles), and a mean field curve (black solid curve). The former corresponds to the onset of superconducting fluctuation terminating at the QCP characterized by the quantum Griffiths singularity at $B = B_c^*$ and $T = 0$ K. The latter $T_{cross}(B)$ represents the crossover boundary between the thermal and quantum vortex creep region derived from the Arrhenius plot analysis of $R_{sheet}(T)$ data in Fig. 1b (see Supplementary Note 5 for details). Because $B_c(T)$ is extrapolated to the onset temperature $T_{onset}$ of the Aslamazov-Larkin type fluctuation[29] at $B = 0$, the region above $T_c^{MF}(B)$ at low magnetic fields is dominated by thermal fluctuation of the order parameter amplitude. This state crossovers to the thermal creep (thermal phase fluctuation) below $T_c^{MF}(B)$ and then to the quantum creep (quantum phase fluctuation) region below $T_{cross}(B)$ with decreasing temperature. At high magnetic fields, on the other hand, thermal fluctuation region shrinks because of the decrease of $T_c^{MF}(B)$. However, with further decreasing temperature, $B_c(T)$ shows an upturn, which is a signature of the emergence of quantum fluctuation. The $T_{cross}(B)$ curve seems smoothly connected to the upturn of $B_c(T)$ at low temperatures. This feature suggests that the quantum fluctuations in highly crystalline 2D superconductors govern the evolution of the ground state, stemming from the quantum metallic state at a very low magnetic field (Fig. 4a) to the quantum Griffiths state (Fig. 4b), followed by the QPT above $B_{c2}^{MF}(0)$ with the quantum Griffiths singularity (see Supplementary Note 6 for details). Such upturn behavior as observed in ZrNCl-EDLT was not observed in $MoS_2$-EDLT. The absence of upturn behavior is possibly because of lower $T_c$ and thus narrower measurement range in $MoS_2$-EDLT compared with the that in ZrNCl-EDLT as discussed in the case of $LaAlO_3/SrTiO_3$[13]. Nevertheless, the signature of divergent behavior of zv observed in the $MoS_2$-EDLT suggests the existence of the quantum Griffiths state.

It is noted that the quantum Griffiths state, which shows activated critical behavior with divergent exponent, is different from the phase separation or originally spatial inhomogeneity of the superconducting state as is discussed in $LaTiO_3/SrTiO_3$, which shows multiple critical exponents[30,31]. In the early works on amorphous 2D superconductors, the quantum Griffiths singularity was not observed even at ultra-low temperature[23]. One possible reason for the absence of quantum Griffiths singularity, in addition to the measurement temperature range[11], is the large amount of homogeneously distributed quenched disorder, in the conventional 2D superconductors[2], which give a rather homogeneous effect. On the other hand, in highly crystalline 2D systems, the ordered regions and weakly disordered regions can coexist, which causes rare regions under sufficiently large magnetic fields, and thus leads to quantum Griffiths singularity. Therefore, the quantum Griffiths state is a signature of high crystallinity of the system.

In the recent study of mechanically exfoliated 2D NbSe$_2$[5], neither the single crossing nor the multiple crossing behavior has been observed, leading to the absence of the quantum Griffiths state. This is possibly due to the further reduction of the quenched disorder in 2D single crystalline NbSe$_2$. In this case, the localization effect at low temperatures might be too small to observe the quantum Griffiths state.

In conclusion, we established the comprehensive quantum phase diagram including superconducting fluctuation regime in ion-gated 2D crystalline superconductors. This class of superconductors commonly exhibit the wide range of quantum metallic state and quantum Griffiths state as a ground state, which transfers into the weakly localized normal metal through a QPT showing the quantum Griffiths singularity. Furthermore, our results indicate that the coexistence of the puddle-like superconducting regions and the dissipation region, occurring owing to the combined effect of the strong quantum fluctuation and extremely weak pinning or disorder, is universal nature above $B_{c2}^{MF}$ in the low temperatures in highly crystalline 2D superconductors.

**Methods**

**Device fabrication.** Bulk ZrNCl and 2*H*-polytype MoS$_2$ single crystals were mechanically exfoliated into thin flakes with dozens of nanometers in thickness using the Scotch-tape method, and then, the flakes were transferred onto a SiO$_2$/Si substrate. Au (90 nm)/Cr (5 nm) electrodes were patterned onto an isolated thin flake in a Hall bar configuration, and a side gate electrode was patterned onto the substrate. A droplet of ionic liquid covered both the channel area and the gate electrode. Ionic liquid N,N-diethyl-N-(2-methoxyethyl)-N- methylammonium bis (trifluoromethylsulphonyl) imide (DEME-TFSI) was selected as a gate medium. The sheet carrier densities $n_{2D}$ induced electrostatically for the ZrNCl-EDLT at $V_G$ = 6.0 and the MoS$_2$-EDLT at $V_G$ = 5.0 V, which were determined by the Hall effect measurements at 30 and 15 K, were $4.8 \times 10^{14}$ cm$^{-2}$ and $1.3 \times 10^{14}$ cm$^{-2}$, respectively.

**Transport measurements.** The temperature dependent resistance under magnetic fields was measured with a standard four-probe geometry in a Quantum Design Physical Property Measurement System (PPMS) combined with two kinds of AC lock-in amplifiers (Stanford Research Systems Model SR830 DSP and Signal Recovery Model 5210). The gate voltage was supplied by a Keithley 2400 source meter at 230 K, which is just above the glass transition temperature of DEME-TFSI, under high vacuum (less than 10$^{-4}$ Torr).

**Data availability**. The data that support the findings of this study are available from the corresponding author upon reasonable request.

**References**

1. Fisher, M. P. A. Quantum phase transitions in disordered two-dimensional superconductors. *Phys. Rev. Lett.* **65,** 923–927 (1990).
2. Goldman, A. M. Superconductor-insulator transitions. *Int. J. Mod. Phys. B* **24,** 4081–4101 (2010).
3. Saito, Y., Nojima, T. & Iwasa, Y. Highly crystalline 2D superconductors. *Nat. Rev. Mater.* **2,** 16094 (2016).
4. Saito, Y., Kasahara, Y., Ye, J., Iwasa, Y. & Nojima, T. Metallic ground state in an ion-gated two-dimensional superconductor. *Science* **350,** 409–413 (2015).
5. Tsen, A. W. *et al.* Nature of the quantum metal in a two-dimensional crystalline superconductor. *Nat. Phys.* **12,** 208–212 (2016).
6. Reyren, N. *et al.* Superconducting interfaces between insulating oxides. *Science* **317,** 1196–1199 (2007).
7. Gozar, A. *et al.* High-temperature interface superconductivity between metallic and insulating copper oxides. *Nature* **455,** 782–785 (2008).
8. Cao, Y. *et al.* Quality heterostructures from two-dimensional crystals unstable in air by their assembly in inert atmosphere. *Nano Lett.* **15,** 4914–4921 (2015).
9. Xi, X. *et al.* Strongly enhanced charge-density-wave order in monolayer $NbSe_2$. *Nat. Nanotechnol.* **10,** 765–769 (2015).
10. Ye, J. T. *et al.* Superconducting dome in a gate-tuned band insulator. *Science* **338,** 1193–1196 (2012).
11. Xing, Y. *et al.* Quantum Griffiths singularity of superconductor-metal transition in Ga thin films. *Science* **350,** 542–545 (2015).
12. Xing, Y. *et al.* Ising superconductivity and quantum phase transition in macro-size monolayer $NbSe_2$. *Nano Lett.* **17,** 6802–6807 (2017).
13. Shen, S. *et al.* Observation of quantum Griffiths singularity and ferromagnetism at the superconducting $LaAlO_3/SrTiO_3$ interface. *Phys. Rev. B* **94,** 144517 (2016).
14. Griffiths, R. B. Non magnetic behavior above the critical point in a rondom Ising ferromagnet. *Phys. Rev. Lett.* **23,** 17–19 (1969).
15. Spivak, B., Oreto, P. & Kivelson, S. A. Theory of quantum metal to superconductor transitions in highly conducting systems. *Phys. Rev. B* **77,** 214523 (2008).
16. Saito, Y. *et al.* Superconductivity protected by spin–valley locking in ion-gated $MoS_2$. *Nat. Phys.* **12,** 144–149 (2016).

17. Yamada, M., Hirahara, T. & Hasegawa, S. Magnetoresistance measurements of a superconducting surface state of In-induced and Pb-induced structures on Si(111). *Phys. Rev. Lett.* **110,** 237001 (2013).

18. Matetskiy, A. V *et al.* Two-dimensional superconductor with a giant Rashba effect: One-atom-layer Tl-Pb compound on Si(111). *Phys. Rev. Lett.* **115,** 147003 (2015).

19. Ullah, S. & Dorsey, A. T. Critical fluctuations in high-temperature superconductors and the Ettingshausen effect. *Phys. Rev. Lett.* **65,** 2066–2069 (1990).

20. Theunissen, M. H. & Kes, P. H. Resistive transitions of thin film superconductors in a magnetic field. *Phys. Rev. B* **55,** 15183–15190 (1997).

21. Werthamer, N. R., Helfand, E. & Hohenberg, P. C. Temperature and Purity Dependence of the Superconducting Critical Field, Hc2. III. Electron Spin and Spin-Orbit Effects. *Phys. Rev.* **147,** 295–302 (1966).

22. Yazdani, A. & Kapitulnik, A. Superconducting-insulating transition in two-dimensional α-MoGe thin films. *Phys. Rev. Lett.* **74,** 3037–3040 (1995).

23. Qin, Y., Vicente, C. L. & Yoon, J. Magnetically induced metallic phase in superconducting tantalum films. *Phys. Rev. B* **73,** 100505 (2006).

24. Vojta, T. & Schmalian, J. Quantum Griffiths effects in itinerant Heisenberg magnets. *Phys. Rev. B* **72,** 45438 (2005).

25. Vojta, T. & Sknepnek, R. Critical points and quenched disorder: From Harris criterion to rare regions and smearing. *Phys. Status Solidi Basic Res.* **241,** 2118–2127 (2004).

26. Neto, A. H. C., Castilla, G. & Jones, B. A. Non-Fermi liquid behavior and Griffiths phase in f-electron compounds. *Phys. Rev. Lett.* **81,** 3531–3534 (1998).

27. de Andrade, M. C. *et al.* Evidence for a common physical description of non-Fermi-liquid behavior in f-electron systems. *Phys. Rev. Lett.* **81,** 5620–5623 (1998).

28. Ubaid-Kassis, S., Vojta, T. & Schroeder, A. Quantum Griffiths phase in the weak itinerant ferromagnetic alloy $N_{1-x}V_x$. *Phys. Rev. Lett.* **104,** 66402 (2010).

29. Aslamasov, L. G. & Larkin, A. I. The influence of fluctuation pairing of electrons on the conductivity of normal metal. *Phys. Lett. A* **26,** 238–239 (1968).

30. Biscaras, J. *et al.* Multiple quantum criticality in a two-dimensional superconductor. *Nat. Mater.* **12,** 542–548 (2013).

31. Scopigno, N. *et al.* Phase Separation from Electron Confinement at Oxide Interfaces. *Phys. Rev. Lett.* **116,** 26804 (2016).

32. Mason, N. & Kapitulnik, A. True superconductivity in a two-dimensional superconducting-insulating system. *Phys. Rev. B* **64,** 60504 (2001).

**Acknowledgements**

We thank K. Kanoda, S. Okuma, S. Kaneko and K. Ienaga for fruitful discussions. Y.S. was supported by the Japan Society for the Promotion of Science (JSPS) through a research fellowship for young scientists (Grant-in-Aid for JSPS Research Fellow, JSPS KAKENHI Grant Number JP15J07681). This work was supported by Grant-in-Aid for Specially Promoted Research (JSPS KAKENHI Grant Number JP25000003) from JSPS and for Scientific Research on Innovative Areas (JSPS KAKENHI Grant Number JP15H05884) from JSPS.

**Author contributions**

All authors discussed the results and wrote the manuscript. Y.S. conceived the idea, designed the experiments, conducted the transport measurements, and analyzed the data.

**Additional information**

Supplementary information is available in the online version of the paper. Reprints and permissions information is available online at www.nature.com/reprints.

Correspondence and requests for materials should be addressed to Y.S.

**Competing financial interests**

The authors declare no competing financial interests.

**Figure Captions**

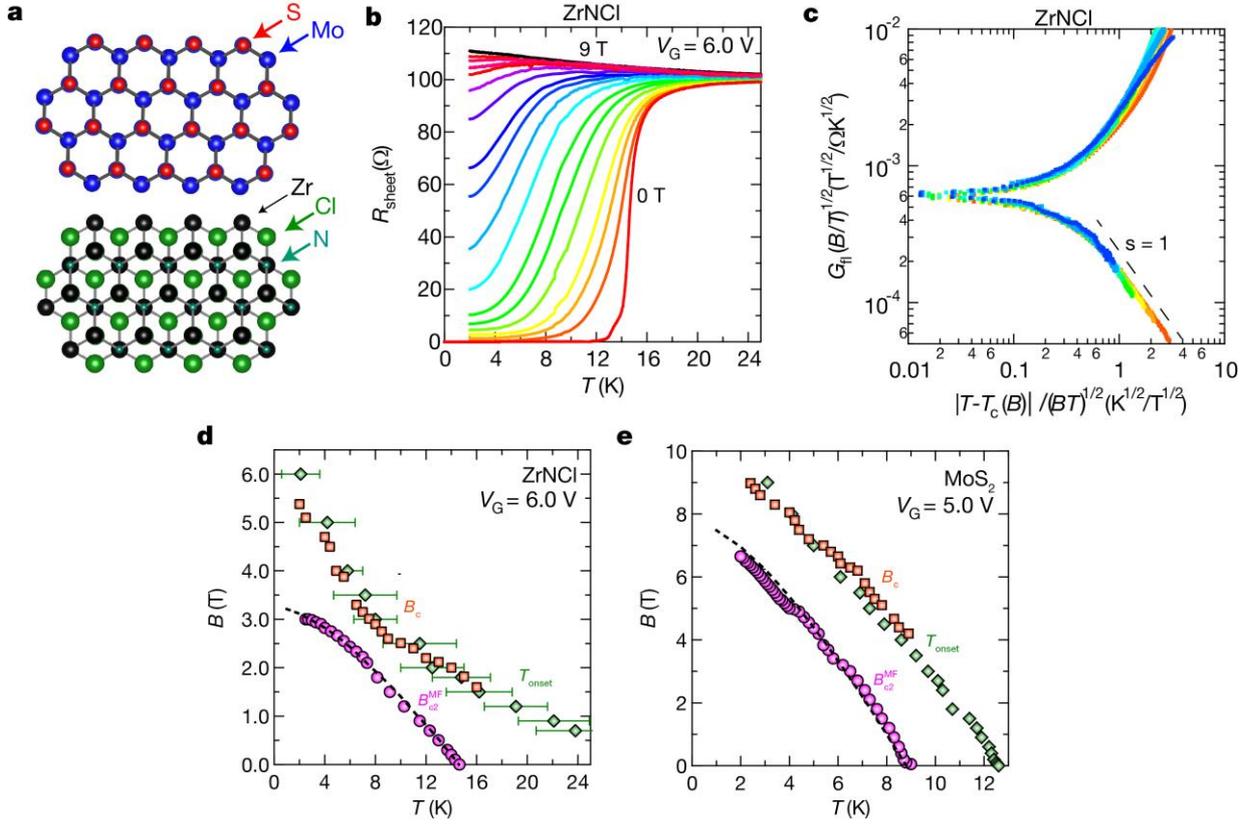

**Figure 1 | Gate-induced superconductivity in ZrNCl and MoS₂ single crystals. a,** Top view of the crystal structures of MoS$_2$ and ZrNCl. **b,** Sheet resistance of a ZrNCl-EDLT as a function of temperature at $V_G$ = 6.0 V, for perpendicular magnetic fields varying in 0.1 T steps from 0 T to 0.3 T, in 0.2 T steps from 0.5 T to 0.9 T, in 0.3 T steps from 1.2 T to 1.8 T, in 0.5 T steps from 2.0 T to 5.0 T, and in 1.0 T steps from 6.0 T to 9.0 T. **c,** Ullah-Dorsey (UD) scaling of fluctuation conductivity using the data of Fig. 1b from 0.1 T to 1.8 T according to equation (1). **d, e,** Superconducting onset and mean-field upper critical field of the ZrNCl- (**d**) and MoS$_2$-EDLT (**e**). Pink circles show the mean field upper critical field $B_{c2}^{MF}$ derived from the UD scaling[19]. Dashed curve shows the fitting by the Wertharmer-Helfand-Hohenberg (WHH) theory[21]. Orange squares show the crossing points $B_c$ of R-B curves in Figs. 2b and d at neighboring temperatures. Green diamonds show the superconducting onset $T_{onset}$. Error bars of green diamonds represent the ambiguity of $T_{onset}$ defined by $dR_{sheet}/dT$ = 0 due to the experimental resolution (see Supplementary Figure 1).

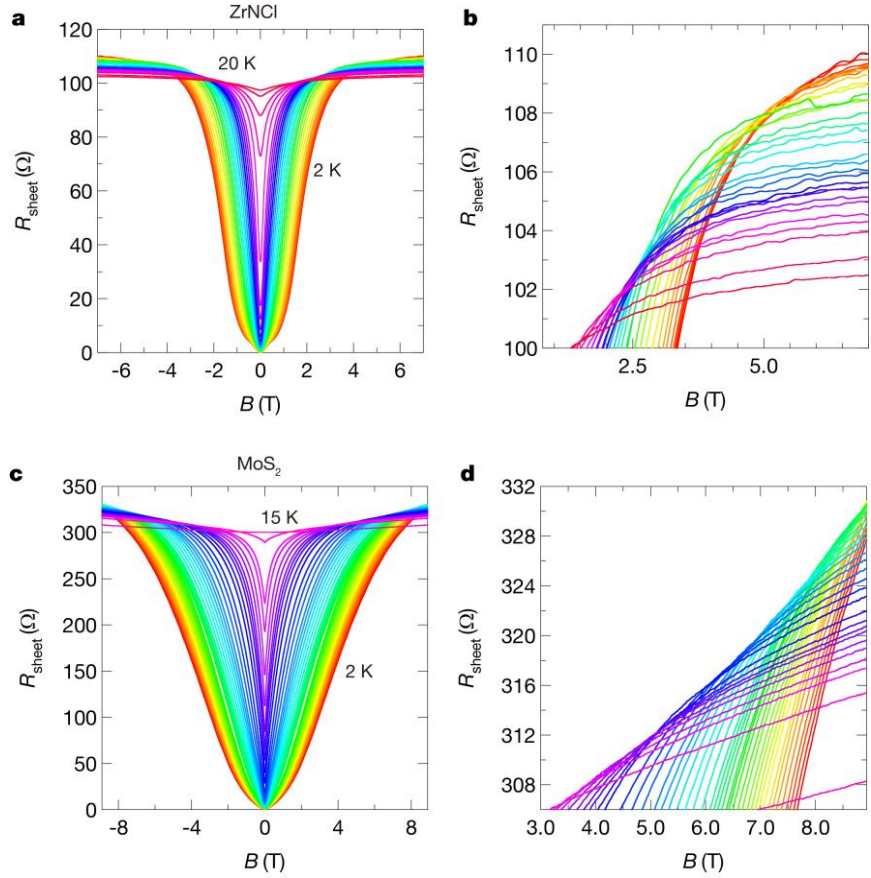

**Figure 2 | Magnetoresistance and multiple critical points in a ZrNCl- and a MoS$_2$-EDLT**. **a**, **c**, Sheet resistance as a function of out-of-plane magnetic field at different temperatures in the ZrNCl-EDLT and the MoS$_2$-EDLT. The measurement temperatures for the ZrNCl-EDLT range from 2 K to 20 K (2, 2.3, 2.5, 2.7, 3, 3.3, 3.7 K, in 0.5 K steps from 4 to 13 K, in 1.0 K steps from 14 to 18 K, and 20 K). The measurement temperatures for the MoS$_2$-EDLT range from 2 K to 15 K (in 0.1 K steps from 2 to 4 K, in 0.2 K steps from 4.2 K to 5.4 K, 6, 6.3, 6.7, 7, 7.3, 7.5, 8, 8.3, 8.7, 9,1 10 and 15 K). **b, d**, A close-up view of around multiple crossing points of *R-B* curves for the ZrNCl-EDLT (**b**) and MoS$_2$-EDLT (**d**).

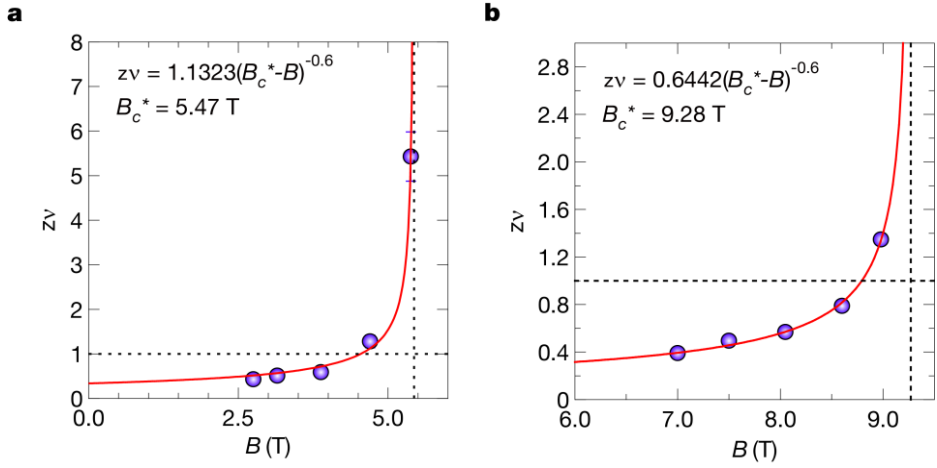

**Figure 3 | Divergent behavior of critical exponent in ion-gated ZrNCl and MoS₂. a, b,** Critical exponent $z\nu$ as a function of magnetic field $B$ for ZrNCl- (**a**) and MoS$_2$-EDLT (**b**). The $z\nu$ values in each temperature regime is calculated by the finite-size scaling (FSS) analysis (see Supplementary Note 4). Error bars represent the width of $z\nu$ value with which the FSS scaling behavior is established in the whole range of $B$ for the data of Supplementary Figure 3. With increasing magnetic field and decreasing temperature, the value of $z\nu$ increases and finally shows divergent behavior with no signature of saturation. The red curve shows a fitting based on the activated scaling law, $z\nu \approx C(B - B_c^*)^{-0.6}$. Two black dashed lines in each graph represent the constant value with $B_c^* = 5.43$ T and $z\nu = 1$ for **a**, respectively, and $B_c^* = 9.27$ T and $z\nu = 1$ for **b**.

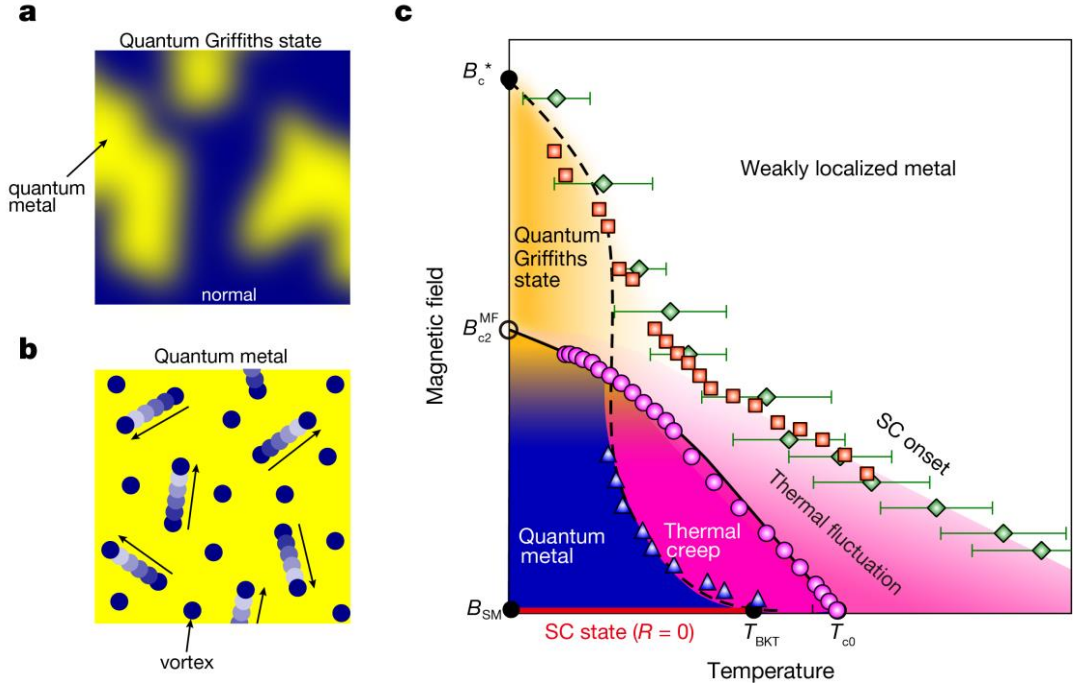

**Figure 4 | $B$-$T$ phase diagram of highly crystalline 2D superconductors. a**, Schematic image of quantum Griffiths state. Blue and yellow regions represent normal and quantum metallic state (rare regions), respectively. **b**, Schematic image of quantum metallic state. In this state, vortices (blue circles) shows creep motion in the superconducting region (yellow region). The quantum metallic phase in Fig. 4a is the same as that in Fig. 4b. **c**, The plotted data based on Supplementary Figure 5a. $T_{c0}$ is the transition temperature determined by the thermal fluctuation theories (Aslamazov-Larkin and Maki-Thompson model), and true superconducting state (zero resistance state) is realized below the Berezinskii-Kosterlitz-Thouless (BKT) transition temperature ($T_{BKT}$)[4]. Blue triangles show the crossover temperature $T_{cross}$ between the thermal creep regime and the quantum creep regime (see Supplementary Figure 4). Orange squares, green diamonds and pink circles show the crossing points $B_c$ of $R$-$B$ curves in Figs. 2b and d at neighboring temperatures, the superconducting onset $T_{onset}$ (see Supplementary Figure 1), and the mean field upper critical field $B_{c2}^{MF}$ derived from the Ullah-Dorsey (UD) scaling[19], respectively. Error bars of green diamonds represent the ambiguity of $T_{onset}$ defined by $dR_{sheet}/dT = 0$ due to the experimental resolution (see Supplementary Figure 1). Under a relatively low magnetic field, the finite resistance state occurs due to the thermal creep (pink region) and quantum creep (blue region) in the relatively high and low temperature region, respectively. $B_{SM}$ is the hypothetical transition magnetic field from the zero-resistance state to the quantum metal[32] (vortex liquid). The system eventually exhibits the quantum Griffiths state (orange region) up to the

characteristic critical magnetic field $B_c^*$. The black solid and dashed curves show the $B_{c2}^{MF}(T)$ based on the Wertharmer-Helfand-Hohenberg (WHH) model and, crossover curve between thermal fluctuation region and quantum fluctuation region, respectively.